\documentclass[prb,a4paper,aps,10pt,twocolumn,floatfix,superscriptaddress,amsfonts]{revtex4}
\pdfoutput=1
\usepackage[T1]{fontenc}\usepackage[latin1]{inputenc}
\usepackage{dcolumn,graphicx,color,booktabs,microtype,afterpage} 
\usepackage{sidecap}
\usepackage{times}
\usepackage[colorlinks,plainpages=false,linkcolor=blue,urlcolor=blue,citecolor=blue,pdfpagemode=UseNone,pdfstartview=FitBH]{hyperref}

\usepackage{amssymb}
\usepackage{amsmath}
\usepackage{graphicx}
\usepackage{esvect}
\usepackage{bm}
\usepackage{tabularx}
\renewcommand{\deg}{^{\circ}}
\newcommand{\ymgo}{YbMgGaO$_4$}
\newcommand{\kervo}{K$_3$Er(VO$_4$)$_2$}

\newcommand{\Seff}{ $S_{\text{eff}} $}
\begin{document}
\preprint{APS/123-QED}
%
%
\title
{Pseudo-Spin Versus Magnetic Dipole Moment Ordering in the Isosceles Triangular Lattice Material \kervo}
\author{Danielle R. Yahne}
\email{Danielle.Harris@colostate.edu}
\affiliation{Department of Physics, Colorado State University, 200 W. Lake St., Fort Collins, CO 80523-1875, USA}
\author{Liurukara D. Sanjeewa}
\affiliation{Materials Science and Technology Division, Oak Ridge National Laboratory, Oak Ridge, Tennessee 37831, United States }
\author{Athena S. Sefat}
\affiliation{Materials Science and Technology Division, Oak Ridge National Laboratory, Oak Ridge, Tennessee 37831, United States }
\author{Bradley S. Stadelman}
\affiliation{Department of Chemistry, Clemson University, Clemson, South Carolina 29634-0973, United States }
\author{Joseph W. Kolis}
\affiliation{Department of Chemistry, Clemson University, Clemson, South Carolina 29634-0973, United States }
\author{Stuart Calder}
\affiliation{Neutron Scattering Division, Oak Ridge National Laboratory, Oak Ridge, Tennessee 37831, United States}
\author{Kate A. Ross}
\email{Kate.Ross@colostate.edu}
\affiliation{Department of Physics, Colorado State University, 200 W. Lake St., Fort Collins, CO 80523-1875, USA}
\date{\today}
%
%
%
%
\begin{abstract}
Spin-$\frac{1}{2}$ antiferromagnetic triangular lattice models are paradigms of geometrical frustration, revealing very different ground states and quantum effects depending on the nature of anisotropies in the model. Due to strong spin orbit coupling and crystal field effects, rare-earth ions can form pseudo-spin-$\frac{1}{2}$ magnetic moments with anisotropic single-ion and exchange properties. Thus, rare-earth based triangular lattices enable the exploration of this interplay between frustration and anisotropy. Here we study one such case, the rare-earth double vanadate glaserite material \kervo, which is a quasi-2D isosceles triangular antiferromagnet. Our specific heat and neutron powder diffraction data from \kervo\ reveal a transition to long range magnetic order at $155 \pm 5$ mK which accounts for all $R\ln2$ entropy. We observe what appears to be a coexistence of 3D and quasi-2D order below $T_N$. The quasi-2D order leads to an anisotropic Warren-like peak profile for $(hk0)$ reflections, while the 3D order is best-described by layers of antiferromagnetic $b$-aligned moments alternating with layers of zero moment. Our magnetic susceptibility data reveal that Er$^{3+}$ takes on a strong XY single-ion anisotropy in \kervo, leading to vanishing moments when \emph{pseudo-spins} are oriented along $c$. Thus, the magnetic structure, when considered from the pseudo-spin point of view comprises alternating layers of $b$-axis and $c$-axis aligned antiferromagnetism.
\end{abstract}
\pacs{xx.xx.mm}
\maketitle
%
%
%
\indent

\section{Introduction}
Magnetic frustration has been of interest in condensed matter physics due to the presence of competing interactions which often leads to exotic properties. A two-dimensional (2D) triangular lattice with antiferromagnetically (AFM) interacting Ising spins is the simplest example of geometrical frustration. Wannier found in 1950 that this model has a macroscopically degenerate ground state and the frustration suppresses order down to zero temperature \cite{Wannier}. A Quantum Spin Liquid (QSL) state, which exhibits quantum entanglement and fractionalized excitations, was first envisioned by Anderson to exist on a 2D triangular Heisenberg AFM (HAFM) \cite{Anderson}. It is now understood that interactions on the 2D triangular HAFM model leads to $120\deg$ order \cite{Bernu_THAFM,Capriotti_THAFM,White_THAFM}, but exchange interaction anisotropies or lattice distortions can lead to other interesting phenomena. For example, the isosceles triangular AFM Cs$_2$CuCl$_4$ was found to be a 1D spin chain and is an example of "dimensional reduction" induced by frustration \cite{Coldea_Cs2CuCl4,Balents}, and anisotropic exchange models on the triangular lattice have been proposed to host QSL phases \cite{Zhu_YMGO,Iaconis,ZhuTopography,Bordelon_NaYbO,Zhong_NaBaCoPO}.

\begin{figure}[!t]
\includegraphics[scale = 1.0]{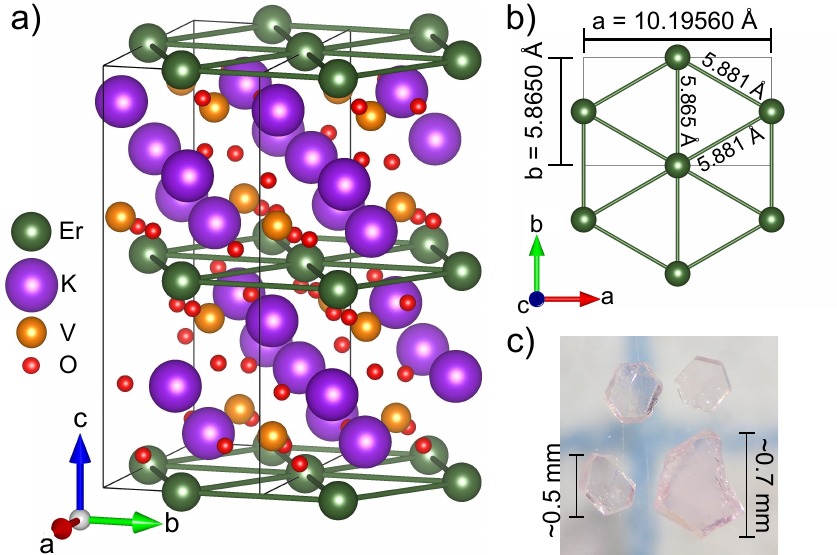}
\caption{ a) Crystal structure of monoclinic \kervo \ (space group $C2/c$) showing layers of 2D isosceles triangular Er$^{3+}$ lattices. b) The 2D isosceles triangular Er$^{3+}$ lattice, with bond lengths and unit cell size shown (not shown, $c = 15.2050$ \AA). c) Typical single crystals of \kervo\, which were co-aligned for magnetization and susceptibility measurements. }
\label{figstructure}
\end{figure}

Rare-earth based frustrated materials have become of interest due to strong spin orbit coupling and crystal electric field (CEF) effects which can lead to \Seff $\ =\frac{1}{2}$ doublets (pseudo-spin-$\frac{1}{2}$) and anisotropic effective exchange models based on these pseudo-spin-$\frac{1}{2}$ moments. This makes them ideal to study quantum phases arising from anisotropic exchange. The relationship between the observed magnetic dipole moments ($\mu_i$) and the pseudo-spin-$\frac{1}{2}$ operators ($S_i$) is given by the $g$-tensor: $\mu_i = g_{ii} S_i$ \footnote{This assumes the $g_{ii}$ values are those obtained from the square root of the eigenvalues of the $G$ tensor \cite{EPR}, so the moment directions defined here by $i$ are along the eigenvectors of that tensor}. Depending on the details of the CEF Hamiltonian, the ground state doublet forming the pseudo-spins can have certain $g$ components become vanishingly small (or in some cases, identically zero due to the symmetry) and thus no appreciable magnetic dipole moment associated with that pseudo-spin direction \cite{Rau_review}. In the case where the symmetry prevents any dipole moment, these pseudo-spin directions are associated with higher multipoles, such as quadrupoles \cite{Onoda,Petit_przro} or octupoles\cite{Huang_QSI,Lhotel_ndzro,Li_DO,Li_cesno}.

In terms of the search for quantum magnetic phases based on rare earth ions, Yb$^{3+}$ has received the most attention. For instance, Yb$_2$Ti$_2$O$_7$, was proposed as a quantum spin ice material \cite{Ross_YbTiO,Applegate_YbTiO,Hayre_YbTiO,Scheie_YbTiO} but was later shown to be an unusual ferromagnet with continuum-like scattering \cite{Chang_higgs,Lhotel_ybtio,Gaudet_Ybtio} that appears to arise from phase competition and non-linear spin wave effects \cite{Robert_ybtio,Thompson_ybtio,Rau_ybtio}. Meanwhile the triangular lattice \ymgo\ was proposed as a QSL but may instead exhibit a random valence bond state due to Mg/Ga site disorder \cite{Li_YMGO,Shen_QSL,Paddison_YMGO,Chen_YMGO,Kimchi_YMGO,Steinhardt_YMGO,Zhu_YMGO}. Frustrated Er$^{3+}$ materials are also of interest, and the pyrochlores (Er$_2$B$_2$O$_7$, B = Ti, Sn, Ge, Pt, etc.) \cite{Zhitomirsky_ertio,Savary_ErTiO,Guitteny_ErSnO,Ross_ErTiO,Li_ErGeO,Dun_ErGeO,Cai_ErPtO,Rau_ertio,Petit_ErSnO,Hallas_ErPtO} have enjoyed the most attention, but other frustrated geometries realized by Er$^{3+}$ are just beginning to be explored \cite{Cai_ermggao,Cai_ergao}. Here we study the isosceles triangular material \kervo\ and show that it has strong XY single ion anisotropy with an unconventional magnetic ground state described by alternating ordered layers of antiferromagnetic "magnetic dipole active" and "magnetic dipole silent" pseudo-spins.

 \kervo \ is a member of the rare-earth double vanadate glaserite family, K$_3$RE(VO$_4$)$_2$, where RE = (Sc, Y, Dy, Ho, Er, Yb, Lu, or Tm). Previous studies on rare-earth double \emph{phosphate} glaserites (K$_3$RE(PO$_4$)$_2$) have shown that there can exist structural transitions between trigonal and lower symmetry structures of these compounds (i.e. monoclinic) \cite{Ushakov_kerpo,Szczygie2008}. While previous reports of \kervo \ describe it in terms of a trigonal space group ($P\overline{3}m1$) at room temperature\cite{Kimani_K3REV2O4}, we have found from powder and single crystal x-ray diffraction, as well as low temperature neutron diffraction, that a monoclinic structure (space group $C2/c$), shown in Fig.~\ref{figstructure}(a) $\&$ (b), is appropriate for our samples at all measured temperatures, similar but not identical to K$_3$Er(PO$_4$)$_2$ (which forms in space group $P2_1/m$).


\section{Experimental Method and Results}
The crystal growth of monoclinic \kervo \ phase involved two steps. First, powder targeting a stoichiometric product of \kervo \ was performed using K$_2$CO$_3$, Er$_2$O$_3$ and (NH$_4$)VO$_3$. A total of $3$ g of components were mixed in a stoichiometric ratio of $3$:$1$:$4$ and ground well using an Agate motor and pestle. The powder mixture was then pressed into pellets and heated to $750\deg$C for $80$ hours. After the reaction period, the resulted pellets were crushed, ground and checked the purity using powder X-ray diffraction (PXRD). According the PXRD, majority phase was matched with the \kervo \ (PDF No. $00$-$51$-$0095$) with impurities of K$_3$VO$_4$ and ErVO$_4$. In the second step, the resulted \kervo \ powder was treated hydrothermally to obtain single crystals.

Hydrothermal synthesis was performed using $2.75$-inch long silver tubing that had an inner diameter of $0.375$ inches. After silver tubes were welded shut on one side, the reactants and the mineralizer were added. Next, the silver ampules were welded shut and placed in a Tuttle-seal autoclave that was filled with water in order to provide appropriate counter pressure. The autoclaves were then heated to $600\deg$C for $14$ days, reaching an average pressure of $1.7$ kbar, utilizing ceramic band heaters. After the reaction period, the heaters were turned off and the autoclave cooled to room temperature. Crystals were recovered by washing with de-ionized water. In a typical reaction $0.4$ g of \kervo \ powder was mixed with a mineralizer solution of $0.8$ mL of $10$ M K$_2$CO$_3$.

Crystals of \kervo , used for magnetism and heat capacity measurements, were physically examined and selected under an optical microscope equipped with a polarizing attachment. Room temperature single crystal structures were characterized using a Bruker D$8$ Venture diffractometer Mo K$\alpha$ radiation ($\lambda = 0.71073$ \r{A}) and a Photon $100$ CMOS detector. The Bruker Apex$3$ software package with SAINT and SADABS routines were used to collect, process, and correct the data for absorption effects. The structures were solved by intrinsic phasing and subsequently refined on $F^2$ using full-matrix least squares techniques by the SHELXTL software package\cite{Sheldrick}. All atoms were refined anisotropically.

\begin{figure}[!t]
\includegraphics[scale = 0.8]{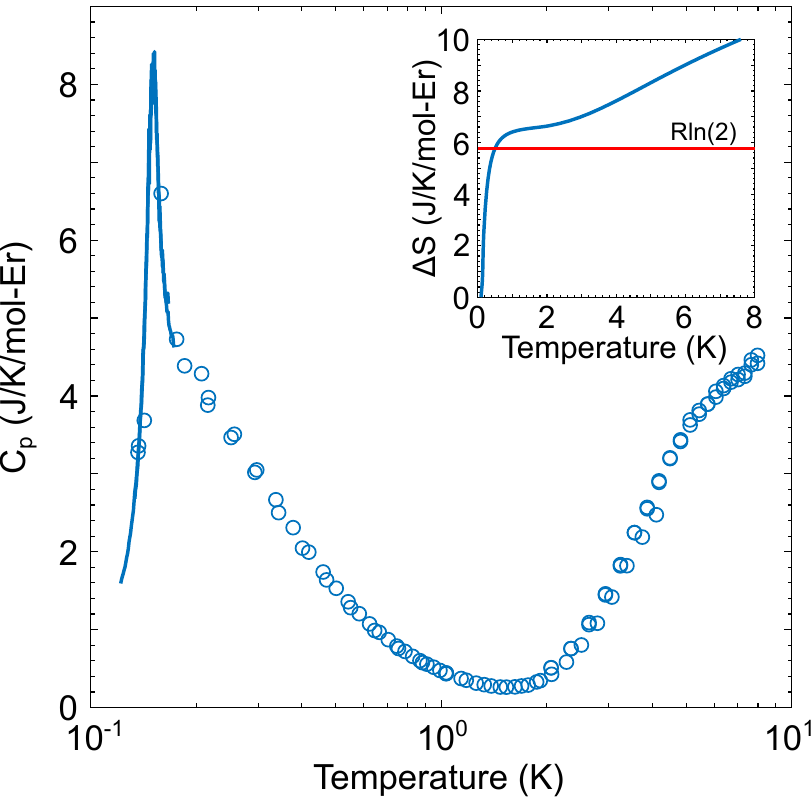}
\caption{ Single crystal heat capacity in zero magnetic field. The open circles represent the adiabatic measurements, while the solid line represents the large $\Delta T$ measurements. A sharp magnetic transition is observed at $155$ mK, with the asymmetric shape on the high temperature side of the transition indicating a build-up of low dimensional short range correlations. (Inset) Entropy calculated from $C_p$ vs. $T$ with the $R\ln{2}$ limit shown, indicating an isolated \Seff \ $=\frac{1}{2}$ system below $1$ K. }
\label{figHC}
\end{figure}

We performed heat capacity measurements from $8$ K down to $50$ mK (Fig.~\ref{figHC}) on a $0.41 \pm 0.05$ mg single crystal sample (examples shown in Fig.~\ref{figstructure}(c)) using a Quantum Design PPMS with dilution refrigerator insert. We employed two measurement techniques, a typical quasi-adiabatic thermal relaxation measurement with temperature rise $\Delta T/T$ of $2\%$, as well as ``long pulse'' measurements where $\Delta T/T$ can be as large as $400\%$, as described in Ref. \onlinecite{Scheie_LP}. We find a sharp transition at $T_N = 155 \pm 5$ mK, much lower than the Curie-Weiss temperature (discussed later), indicating that this system is frustrated as expected, with a frustration parameter of $f = \theta_{CW}/T_N \approx 20$. The total $C_p(T)$ (not lattice subtracted) reveals a broad peak around $10$ K, the shape of which cannot be purely attributed to a power law contribution from acoustic phonon modes, as well as a gradual release of entropy on cooling from $1$ K down to $155$ mK, at which temperature a sharp anomaly is observed. The high temperature peak near $10$ K is indicative of a low-lying excited CEF multiplet with energy of about $2$ meV. The entropy change between $50$ mK to $2$ K accounts for all the entropy expected from a Kramers CEF ground state doublet ($R\ln{2}$ per mole Er$^{3+}$, see inset of Fig.~\ref{figHC}). Less than $30\%$ of this entropy is released via the sharp anomaly, indicating that short range correlations develop over a broad temperature range above the ordering transition. This is commonly found in low dimensional and frustrated magnets, where ordering is suppressed but is eventually triggered by some subleading energy scale in the Hamiltonian (such as inter-layer interactions in the case of quasi-2D systems) \cite{deJongh}. The quasi-2D nature of the magnetism in \kervo\ is also confirmed by neutron powder diffraction to coexist with 3D order below $T_N$, as discussed later.

\begin{figure}[!t]
\includegraphics[scale = 1.0]{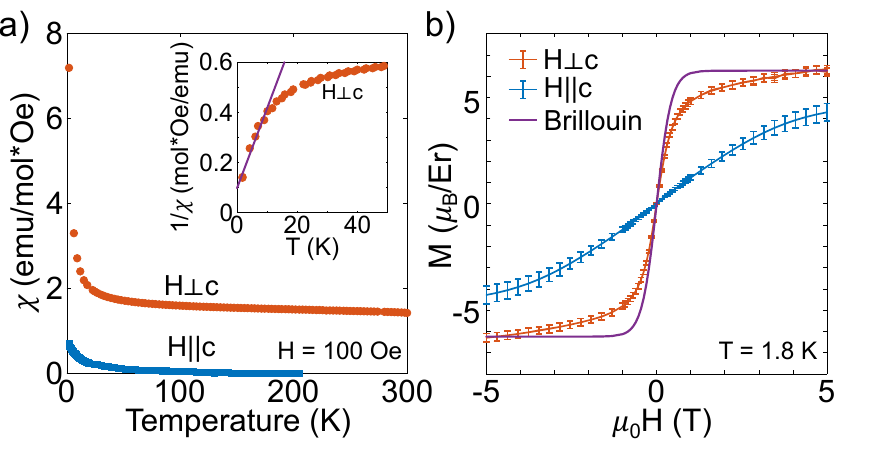}
\caption{ a) Magnetic susceptibility of co-aligned single crystals with the field aligned perpendicular and parallel to the $c$-axis, showing $\chi_{\bot c}$ is larger than $\chi_{|| c}$ by a factor of $\sim 10$. (Inset) Low temperature fit of inverse susceptibility used to find the $H\bot c$ Curie-Weiss temperature $\theta_{CW} \approx -3$ K. b) Magnetization of co-aligned single crystals at $1.8$ K. For $H||c$, significant field-induced mixing of the next highest CEF level produces an enhanced moment in the field. }
\label{figmagnetization}
\end{figure}

The temperature dependent magnetic susceptibility of \kervo \ was measured down to $1.8$ K in a $100$ Oe field (Fig.~\ref{figmagnetization}(a)) using the MPMS XL Quantum Design SQUID magnetometer on $1.60 \pm 0.05$ mg and $1.04 \pm 0.10$ mg of co-aligned single crystals, aligned in the $H\bot c$ and $H||c$ directions respectively. For magnetic fields $H\bot c$, we find net antiferromagnetic interactions shown by the negative Curie-Weiss temperature $\theta_{CW} \approx -3 \ \text{K}$ obtained by fitting between $2$ and $10$ K (although this value is highly dependent on the exact fitting range used due to crystal field effects), similar to \ymgo with $\ \theta_{CW} \approx -4 \ \text{K}$. The magnetic susceptibility $\chi_{||c}$ is an order of magnitude less than $\chi_{\bot c}$, indicating a strong XY nature of the $g$-tensor of Er$^{3+}$ in this material. Magnetization measurements (Fig.~\ref{figmagnetization}(b)), taken at $1.8$ K in a magnetic field up to $5$ T, corroborate that \kervo \ is a strongly XY system due to the large saturation magnetization for $M_{\bot c}$. Neither $M_{\bot c}$ nor $M_{||c}$ follow a Brillouin function expected for a simple paramagnet, suggesting that there is significant mixing of the higher CEF states causing the response to be non-paramagnetic. Due to field induced mixing of the excited CEF levels, the saturation magnetization is not a good indicator of the zero-field $g$-tensor for either direction. Regardless of the CEF mixing, the magnetization $M_{||c}$ starts with a low $g$-value near zero field, consistent with a small $g$-value in the $c$-axis.

Neutron powder diffraction was performed on the HB-2A beamline at Oak Ridge National Laboratory's (ORNL) High Flux Isotope Reactor (HFIR). Approximately $2.5$ g of crystals were ground into a fine powder, placed into a copper sample can and filled with $10$ atm of He gas at room temperature, a technique shown to enable sample thermalization of loose powders below $1$ K \cite{Ryan_he3}. Diffraction patterns were obtained from $10$ K down to $50$ mK, with collimator settings open-open-$12$', and a Ge(113) monochromator provided an incident wavelength of $\lambda = 2.41$ \AA. The patterns were collected over a Q-range of $0.18$ \r{A}$^{-1}$ $< Q < 4.64$ \r{A}$^{-1}$ ($4\deg < 2\theta < 126\deg $) with count times of $2$ hours per scan.

\begin{figure}[t]
\includegraphics[scale = 1.0]{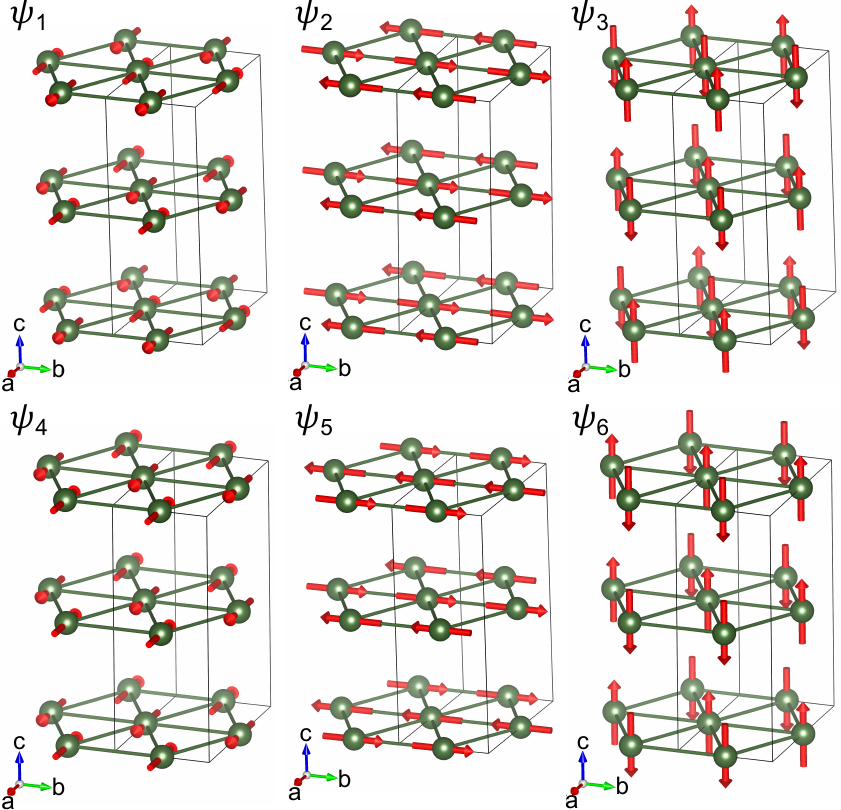}
\caption{ Visualizations of the basis vectors for space group $C2/c$ with k=(1,0,0): $\psi_{1,2,3}$ from $\Gamma_1$ and $\psi_{4,5,6}$ from $\Gamma_3$. All basis vectors are antiferromagnetic in the $ab$-plane. Basis vectors $\psi_{1,3,5}$ are also antiferromagnetic along $c$, while basis vectors $\psi_{2,4,6}$ are ferromagnetic along $c$ (each layer is identical).}
\label{figBVs}
\end{figure}

\begin{figure*}[ht!]
\includegraphics[scale = 1.05]{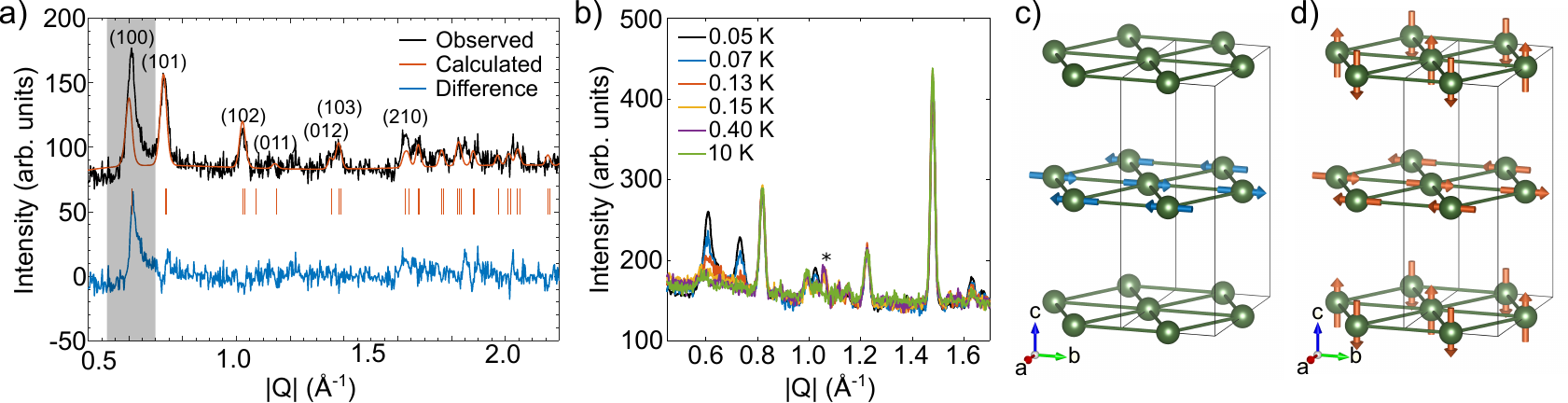}
\caption{ a) Neutron diffraction pattern (black) taken on a $2.5$ g powder sample on beamline HB-2A at HFIR (ORNL). Magnetic peaks were found from subtracting the $400$ mK pattern from the $50$ mK pattern to remove impurity signals. A coexistence of 2D and 3D order was found, and thus the Warren-like $(100)$ peak was not included in the fit. The fit to the 3D order (red) and calculated difference (blue) is shown for the best fit using the Fullprof software, which was a linear combination of equal contributions of basis vectors $\psi_2$ (from $\Gamma_1$) and $\psi_5$ (from $\Gamma_3$).  b) Temperature dependence of magnetic Bragg peaks below the transition temperature of $155$ mK shows the onset of 2D and 3D order occurs at the same temperature. Magnetic impurity peaks were found between $10$ K and $400$ mK, denoted with a star. c) The magnetic structure found from neutron diffraction pattern shows layers of $b$-aligned moment alternating with layers of zero moment, proposed to be due to the strong XY nature of \kervo \ ($g_z \sim 0$). d) The proposed \emph{pseudo-spin} structure, alternating between layers of $b$-aligned pseudo-spin and layers of $c$-aligned pseudo-spin. }
\label{figfullprof}
\end{figure*}

Analysis of the powder diffraction data was performed using the FullProf software suite which implements the Rietveld refinement method \cite{Fullprof_Rodriguez}. The $10$ K data was used to refine the nuclear structure with contributions from the copper cell and aluminum windows masked. Magnetic peaks which could not be indexed within the \kervo\ unit cell emerged between $10$ K and $400$ mK, indicating the presence of magnetic impurities in the sample, which were unable to be identified. These impurities are likely to be from by-products produced during the crystal synthesis, which are easy to avoid for single crystals measurements, but is impractical to completely remove for the large sample mass needed for neutron powder measurements. To remove the impurity signal from the magnetic structure analysis, we subtracted the $400$ mK data from the $50$ mK data, leaving only contributions from \kervo\ magnetic Bragg peaks (Fig.~\ref{figfullprof}(a)). The magnetic peaks indexed gave an ordering wavevector of k=(1,0,0), for which the decomposition of the magnetic representation into irreducible representations (IR's) is $\Gamma_{mag} = 3\Gamma_1^{(1)} + 0\Gamma_2^{(1)} + 3\Gamma_3^{(1)} + 0\Gamma_4^{(1)}$ for a magnetic atom at site $(0.5,0,0.5)$ found using the SARA\textit{h} Representational Analysis software\cite{Sarah_wills} (Kovalev tables). $\Gamma_1$ is composed of basis vectors $\psi_{1,2,3}$, and $\Gamma_3$ is composed of basis vectors $\psi_{4,5,6}$. Basis vectors $\psi_{2,4,6}$ have antiferromagnetic (AFM) spin arrangements in the $ab$ plane which are ferromagnetically (FM) correlated along the $c$-axis (i.e. every layer is identical), with moments pointing along the $b, a,$ and $c$ axes, respectively. $\psi_{1,3,5}$ are AFM in the $ab$ plane as well as along the $c$-axis, with moments pointing along the $a, c,$ and $b$ axes, respectively. The summary of these basis vectors and their components for each site is in Table~\ref{sarah} and shown in Fig.~\ref{figBVs}.

\begin{table}[]
\begin{tabularx}{0.45\textwidth} { 
   >{\centering\arraybackslash}X 
   >{\centering\arraybackslash}X 
   >{\centering\arraybackslash}X
   >{\centering\arraybackslash}X
   >{\centering\arraybackslash}X
   >{\centering\arraybackslash}X
   >{\centering\arraybackslash}X
   >{\centering\arraybackslash}X  }
\toprule
IR         & BV       & \multicolumn{6}{l}{Basis Vector Components}                     \\ \cline{3-8}
           &          & m$_{1a}$ & m$_{1b}$ & m$_{1c}$ & m$_{2a}$ & m$_{2b}$ & m$_{2c}$ \\
\midrule
$\Gamma_1$ & $\psi_1$ & 2        & 0        & 0        & -2       & 0        & 0        \\
           & $\psi_2$ & 0        & 2        & 0        & 0        & 2        & 0        \\
           & $\psi_3$ & 0        & 0        & 2        & 0        & 0        & -2       \\

$\Gamma_3$ & $\psi_4$ & 2        & 0        & 0        & 2        & 0        & 0        \\
           & $\psi_5$ & 0        & 2        & 0        & 0        & -2       & 0        \\
           & $\psi_6$ & 0        & 0        & 2        & 0        & 0        & 2   \\    
\bottomrule
\end{tabularx}
\caption{Irreducible representation and basis vector composition for space group $C2/c$ with k=(1,0,0) found using the SARA\textit{h} Representational Analysis software. The atoms are defined according to m$_1$: $(0.5,0,0.5)$ and m$_2$:$(0.5,0,0)$.}
\label{sarah}
\end{table}

We attempted to fit the magnetic scattering within a single IR, which would be expected for a second order phase transition \cite{GroupTheory_Wills}, however, no linear combination of the basis vectors restricted to a single IR came close to reproducing the observed magnetic structure (Appendix~\ref{appendix:magstruc}). It should also be noted that all fits lacked perfect agreement with the intensity of all of the magnetic peaks simultaneously, specifically with respect to the $(100)$ reflection. The shape of the $(100)$ peak does not follow the typical pseudo-Voigt peak shape, and is instead reminiscent of the Warren line-shape for random 2D layered lattices \cite{Warren}, having an asymmetric base that extends further to high $Q$. In a 2D random layer lattice, where no correlations exist between layers, the structure factor for $(hkl)$ zone centers is expected to be zero \cite{Warren}, in contrast to the $(hk0)$ zone centers, which are non-zero and will have this asymmetric shape. For \emph{quasi-random} 2D layers with some short range correlations between planes, intensity is expected at $(hkl)$ reflections, but peaks will have suppressed intensities and will in principle be broadened compared to a peak arising from long range 3D order\cite{Pappis1961,Shi1993,Fujimoto2003}.

As a pure Warren line-shape did not accurately reproduce the $(100)$ reflection, we explored the 2D nature of this material by performing a numerical simulation for the $(100)$ reflection to determine the in-plane and out-of-plane correlation lengths (see Appendix~\ref{appendix:sim} for details). The simulation of the $(100)$ peak produces an out-of-plane correlation length that is inconsistent with the $(hkl)$ magnetic Bragg peaks, which are almost resolution-limited, indicating there is instead a coexistence of 2D and 3D order in \kervo. The origin of this coexistence is unknown, but similar effects have observed in other materials and is speculated to be caused by structural inhomogeneities \cite{Garlea2011, Garlea2018}. However, as shown in Fig.~\ref{figfullprof}(b), the 2D and 3D peaks have similar temperature dependence, which indicates that even if there are inhomogeneous regions of 3D and quasi-2D order, they onset at the same temperature.

Due to the contributions from 2D and 3D correlations to the $(100)$ peak, it was excluded from the fit of the 3D magnetic structure. Our refined magnetic structure is given by equal contributions from basis vectors $\psi_2$ (from $\Gamma_1$) and $\psi_5$ (from $\Gamma_3$), with moments along $b$ that add together in one layer and cancel in the other layer due to the FM and AFM spin correlations along the $c$-axis (Fig.~\ref{figfullprof}(c)). It should be noted that less prominent contributions of basis vector $\psi_6$, which adds small $c$-aligned moment to the layers, could be included without affecting the fit drastically. From the susceptibility data though, little to no moment is expected out-of-plane, so the $\psi_{6}$ contribution is expected to be small or zero.  Comparing the calculated diffraction pattern to the data (Fig.~\ref{figfullprof}(a)), it is clear that the ($hk0$) peaks are under-estimated.  This is as expected, since the ($hk0$) peaks contain significant contributions from the 2D correlations in the material that are not captured by the model.


\section{Discussion}

The refined magnetic structure, in conjunction with the heat capacity, which produces the full $R\ln{2}$ entropy change upon integrating $C_p/T$ from $50$ mK to $1$ K, suggests that \kervo \ is in a fully ordered state, yet the refined structure implies the absence of ordered moments every other layer.  Quantum fluctuations could in principle produce a reduced or zero moment on some layers, however, a simpler explanation seems to be possible by considering the inferred $g$-tensor and the likely pseudo-spin order. We suggest that the  \emph{pseudo-spin} ordering structure involves the 2D triangular layers alternating between AFM ordered layers with moment along $\vv{b}$ and $\vv{c}$ (Fig.~\ref{figfullprof}(d)) \footnote{this \emph{pseudo-spin} state can be visualized using a combination of $\Psi_{2,5,6}$ and $-\Psi_{3}$ (Fig.~\ref{figBVs})}. Such a spin structure is not likely to be obtained from purely XY exchange interactions. Yet, because of the strong XY \emph{single-ion} nature of this material ($g_z \sim 0$), the layers with the pseudo-spins pointing along the $c$-axis would carry approximately zero dipole moment and thus appear to be disordered (or strongly reduced) according to probes that are sensitive only to dipole magnetic moments, such as neutron scattering. This result emphasizes a subtle point which is sometimes misunderstood; the $g$-tensor anisotropy of pseudo-spin-$\frac{1}{2}$ systems does not need to be the same as the exchange anisotropy.

Similar effects are at play in some rare earth pyrochlores, where the the XY part of the pseudo-spin carries a quadrupolar \cite{Onoda,Petit_przro} or octupolar \cite{Huang_QSI,Li_cesno,Lhotel_ndzro,Li_DO} moment but no dipole moment.  However, even "conventional" Kramer's doublets which transform as dipoles in all directions can have very small $g$-values in certain directions, which is the case for Er$^{3+}$ in Er$_2$Sn$_2$O$_7$ \cite{Guitteny_ErSnO}. Due to the low point symmetry at the Er$^{3+}$ site (triclinic) of \kervo, the ground state CEF doublet is most likely to be a conventional Kramer's doublet.  This could in principle be investigated by an analysis of the CEF levels in the material, however we note that the point symmetry for Er$^{3+}$ in \kervo\ is triclinic, leading to 15 symmetry-allowed Steven's parameters which are unlikely to be determined uniquely by experiment or calculation.


\section{Conclusions}

We have performed an extensive  study of the magnetic properties of a member of the rare-earth double vanadate glaserite materials, which form 2D isosceles (or equilateral, in the case of the trigonal polymorphs) triangular lattices. We found an antiferromagnetic transition in \kervo\ at $155$ mK despite a relatively strong AFM interaction of $3$ K inferred from Curie-Weiss analysis (frustration parameter $f \sim 20$). Susceptibility measurements reveal this material to have strong XY $g$-tensor anisotropy, although field-induced coupling to a low-lying CEF level near $\sim 2$ meV (inferred from $C_p(T)$) hinders a quantitative estimate of the $g$-tensor via magnetization. The magnetic structure is comprised of large AFM magnetic dipole moments ordering along the $b$ axis direction in every other layer, and magnetic dipole suppressed pseudo-spin order along $c$ in the other layers. \kervo\ thus appears to be one of the clearest examples in which pseudo-spin order results in zero dipole moments. Inelastic neutron scattering studies of \kervo\ could help to validate this model, and could also reveal the inferred low lying CEF level. Further studies of other rare earth glaserites, particularly in their trigonal structural polymorphs, would be intriguing, as they could be promising materials for discovering new quantum magnetic phases due to their pseudo-spin-$\frac{1}{2}$ angular momentum and strong frustration.

\begin{acknowledgements}
We thank Gang Chen and Ovidiu Garlea for useful discussions. This research used resources at the High Flux Isotope Reactor, a DOE Office of Science User Facility operated by the Oak Ridge National Laboratory. Work performed on synthesis, crystal growth, and x-ray diffraction at Clemson University was funded by DOE BES Grant No. DE-SC$0014271$. DRY, KAR, and JWK acknowledge funding from the Department of Energy award DE-SC$0020071$ during the preparation of this manuscript.
\end{acknowledgements}


\appendix

\section{Sample Preparation}
\label{appendix:sampleprep}

\begin{table}[b]
\begin{tabular}{|c|c|}
\hline
Empirical formula                           & \kervo                    \\
\hline
Formula weight (g/mol)                      & $514.44$                  \\
\hline
Crystal system                              & monoclinic                \\
\hline
Crystal dimensions, mm                      & $0.10$ x $0.02$ x $0.02$  \\
\hline
space group, Z                              & $C2/c$ (no.$15$), $4$     \\
\hline
T, K                                        & $298$                     \\
\hline
a, \r{A}                                    & $10.1956(4)$              \\
\hline
b, \r{A}                                    & $5.8650(2)$               \\
\hline
c, \r{A}                                    & $15.2050(6)$              \\
\hline
$\beta, \deg$                               & $90.12(1)$                \\
\hline
Volume, \r{A}$^3$                           & $909.21(6)$               \\
\hline
D(calc), g/cm$^3$                           & $3.758$                   \\
\hline
$\mu$ (Mo K$\alpha$), mm$^{-1}$             & $12.543$                  \\
\hline
F($000$)                                    & $940$                     \\
\hline
T$_{\text{max}}$, T$_{\text{min}}$          & $1.0000, 0.8169$          \\
\hline
2$\theta$ range                             & $2.679-24.990$            \\
\hline
reflections collected                       & $10581 $                  \\
\hline
data/restraints/parameters                  & $781/0/67 $               \\
\hline
final R [$I>2\sigma(I)$] R$_1$, R$_{w2}$    & $0.0372, 0.1144$          \\
\hline
final R (all data) R$_1$, R$_{w2}$          & $0.0374, 0.1144 $         \\
\hline
GoF                                         & $1.086$                   \\
\hline
largest diff. peak/hole, e/ \r{A}$^3$       & $1.760/-1.155$            \\
\hline
\end{tabular}
\caption{ Crystallographic data of \kervo \ determined by single crystal X-ray diffraction. }
\label{table_sc_xrd}
\end{table}

Crystallographic data for monoclinic \kervo \ was determined using single crystal x-ray diffraction, the details of which are outlined in the main text, and the results are shown in Table~\ref{table_sc_xrd}. A large number of single crystals of \kervo \ were ground into a powder for neutron diffraction using a motor and pestle. Due to the large number of crystals necessary to achieve a substantial mass for neutron scattering, the crystals were ground in three batches, which were x-rayed separately and then again after the batches were combined. Powder X-ray diffraction was performed on a Bruker D$8$ Discover Davinci diffractometer from $10\deg < 2\theta < 60\deg$ for approximately $1$ second per $0.02\deg$. The PXRD pattern was fit using TOPAS Reitveld refinement, and was found to be in agreement with the single crystal XRD, as well as no preferred orientation or peak broadening were found, indicating the crystals were ground sufficiently. Impurity peaks were unable to be matched with any of the expected by-products (ErVO$_4$, Er$_3$O$_2$, etc), and are likely to be from by-products introduced during the hydrothermal synthesis. The powder was then shipped to ORNL where it was placed into a copper sample can. The sample can contains a piece of indium within the He filling line to allow the can to be filled with $10$ atm of He gas and then crimped at the indium, thereby containing the gas.

\section{Low Temperature Nuclear Structure}
\label{appendix:nucstruc}

\begin{figure}[t]
\includegraphics[scale = 1.0]{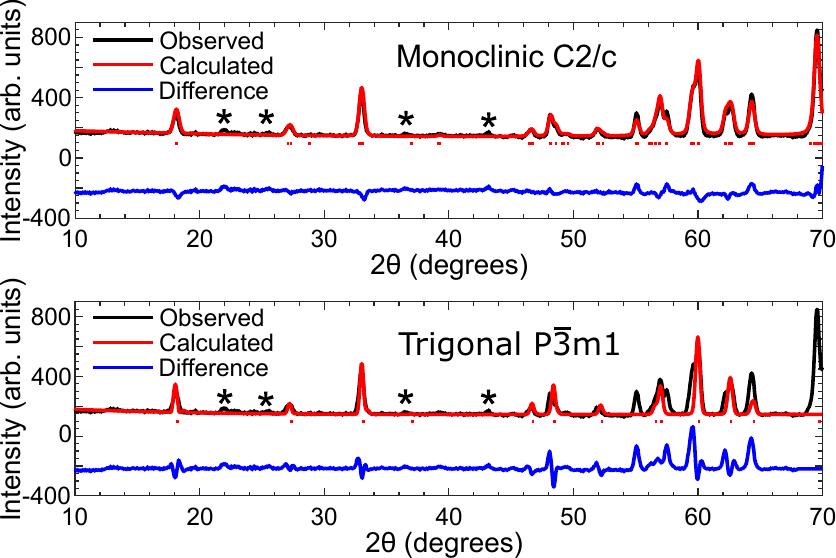}
\caption{ Nuclear structure of \kervo \ found with neutron scattering at 10 K. The trigonal structure clearly does not describe the crystal structure in contrast to the monoclinic structure. The impurity is also seen in small peaks unable to be fit by either structure, denoted by stars. }
\label{figNeutronNuc}
\end{figure}

Neutron powder diffraction data was performed at $10$ K, which was used to determine the low temperature lattice parameters. The neutron data corroborates the monoclinic space group best describes the crystal structure (Fig~\ref{figNeutronNuc}). As expected, we find small impurity peaks in the nuclear data, denoted by stars, and do not find any evidence of preferred orientation. Upon decreasing the temperature to $400$ mK, \emph{magnetic} impurities were found, specifically evidenced by a peak at |Q| = $1.06$ \AA\ that increases in intensity between $10$ K and $400$ mK, and does not increase intensity further upon cooling to $50$ mK (Fig~\ref{figfullprof}(b)). Due to the different onset temperature and no signatures of a transition between $10$ K and $400$ mK in the heat capacity data, this is not believed to be a secondary phase of \kervo. To remove this unknown impurity, we subtracted the $400$ mK data from the $50$ mK data, leaving only the magnetic scattering signal to be analyzed. 

\section{Quasi-2D Simulation}
\label{appendix:sim}

\begin{figure}[b]
\includegraphics[scale = 0.98]{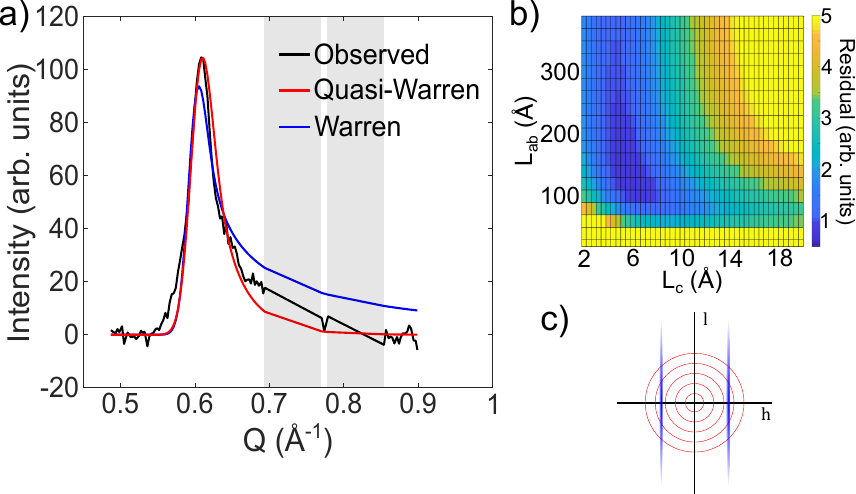}
\caption{ a) Warren fit and quasi-Warren simulation compared with the $50$ mK data (not background subtracted). The $(hkl)$ and nuclear peaks were masked (grey boxes). The Warren fit overestimates the high-Q side of the asymmetric peak. b) Residual plot for varying combinations of correlation lengths along the $ab$ and $c$ directions. c) 2D visualization of simulation in the $h0l$ plane, where blue ellipses are the scattering intensities and the red circles are the integrated areas simulating the powder diffraction.}
\label{figsim}
\end{figure}

The first magnetic peak, $(100)$, did not have the expected pseudo-Voigt peak shape and instead follows more of Warren line-shape for random 2D layer lattices. The Warren line-shape comes from rods of scattering in reciprocal space, centered at $(hk)$. Initially, we attempted to fit the $(100)$ peak with a Warren function \cite{Warren, Wills2000}, but the Warren fit over-estimated the high Q tail (Fig~\ref{figsim}(a)). In addition, if the 2D layers were random with no correlations along the $c$-axis, only $(hk0)$ peaks would have a non-zero structure factor in contrast to $(hkl)$ peaks which would have zero intensity. This suggests that the layers could have some short-range correlations along $c$, thus would be \emph{quasi}-random. Quasi-random 2D layers would still posses asymmetrical $(hk0)$ peaks, while $(hkl)$ peaks would be suppressed and broadened but non-zero. We used a numerical simulation to estimate the in-plane and out-of-plane correlations and fit the $(100)$ asymmetric peak, which shows that this is \emph{not} the case, as discussed next. Thus we infer that the magnetic correlations are a possibly inhomogeneous mixture of 2D and 3D order. 

The numerical simulation was performed by creating a 3D Gaussian ellipse (instead of rods) at zone centers in reciprocal space using the unit cell parameters found from the $10$ K neutron diffraction (Fig~\ref{figsim}(c)). The variables of this ellipse were the standard deviations in the $ab$ and $c$ directions which are related to the correlations in the $ab$ plane, $L_{ab}$, and the correlations between planes, $L_c$, respectively by the equation $L = \sqrt{2\ln{2}}/\sigma$. A radial integral was performed to simulate the powder averaged neutron diffraction pattern, the result of which was then scaled by a Lorentz factor (geometrical correction) and the magnetic form factor. This was then convolved with the instrument resolution, estimated by the FWHM of a nearby nuclear peak. The peak height was scaled to match the data since the intensity is arbitrary. The results compared with the Warren fit are shown in Fig~\ref{figsim}(a). 

The simulation finds a range of correlation lengths fit the data well (Fig~\ref{figsim}(b)), but a best fit estimates correlation lengths along the $c$ axis $L_c \approx 6$ \AA\ (approximately half a unit cell), while correlations in the $ab$ plane $L_{ab} \approx 120$ \AA\ (more than 20 unit cells). When these correlation lengths are applied to an $(hkl)$ peak, we find the peak would be much broader and significantly more suppressed than what we observe. Thus, the 2D and 3D order must be coexisting and onset at the same temperature.

\begin{figure}[ht]
\includegraphics[scale = 1.0]{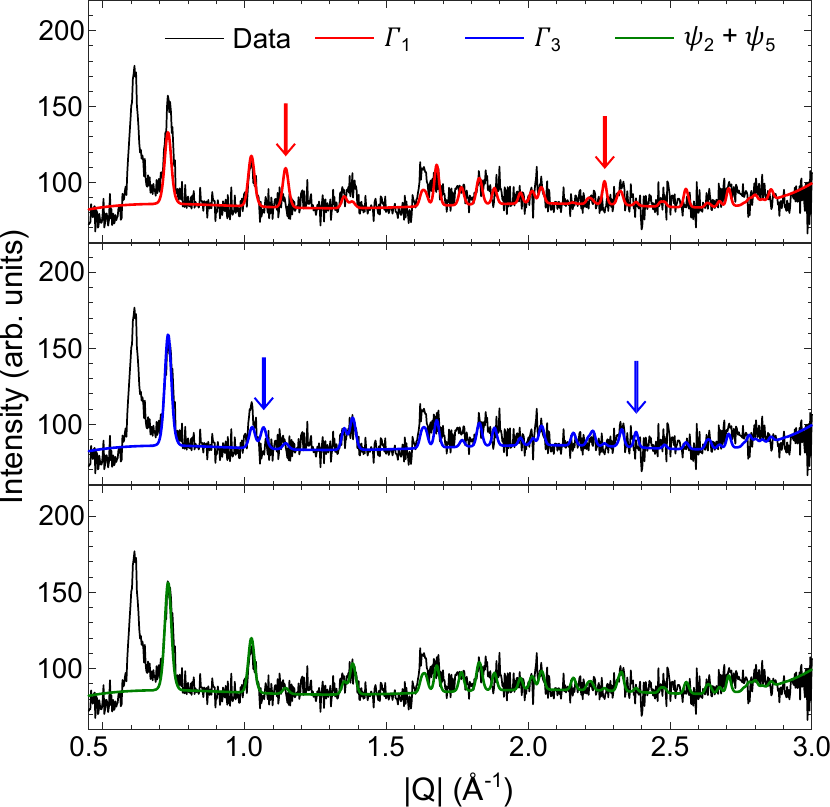}
\caption{ Examples of fits using a single irreducible representation. Both $\Gamma_1$ and $\Gamma_3$ fits give peaks not seen in the data, shown by arrows. Therefore, the best fit comes from a linear combination of $\Gamma_1$ and $\Gamma_3$'s basis vectors, $\psi_2$ and $\psi_5$ respectively.}
\label{figSingleIR}
\end{figure}

\section{Magnetic Structure}
\label{appendix:magstruc}

\begin{figure}[hb]
\includegraphics[scale = 0.6]{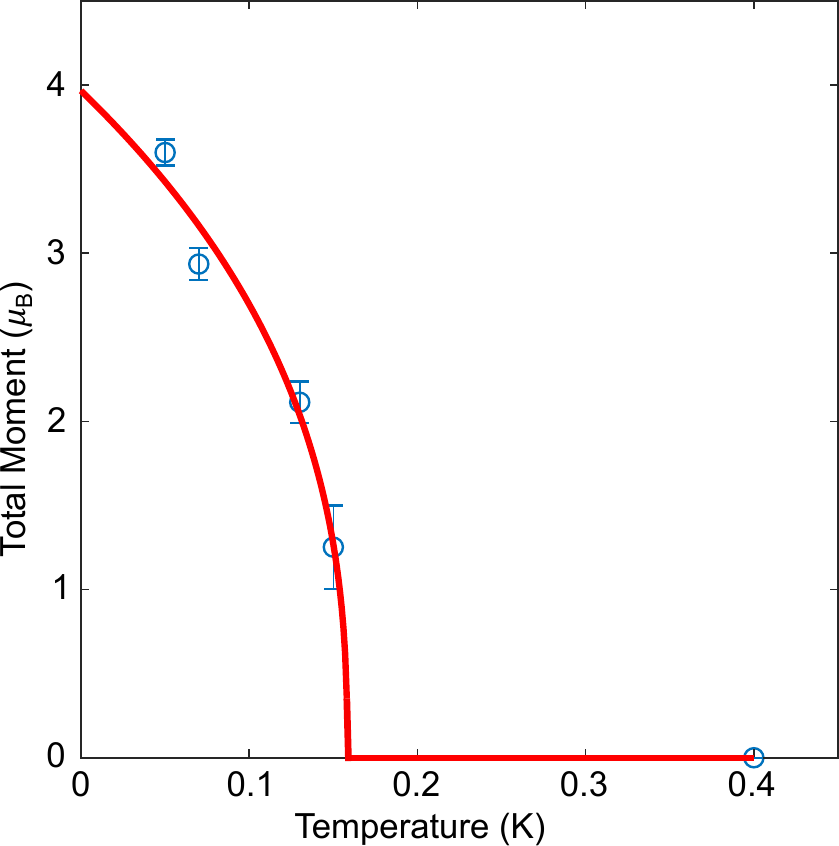}
\caption{ Total moment as a function of temperature. Red line serves as an order parameter guide to the eye, note this is not a fit. }
\label{figMomentvsTemp}
\end{figure}

We attempted to fit the magnetic structure using a single IR, as it was not clear if the transition found in heat capacity was a first- or second-order transition. Examples of those fits are shown in Fig~\ref{figSingleIR}. Both fits of individual IR's had peaks which were not seen in the scattering signal, while the accepted fit (combination of $\psi_2$ from $\Gamma_1$ and $\psi_5$ from $\Gamma_3)$) does not show any spurious peaks. In the scenario where the magnetic structure is a combination of more than one IR, it follows that the transition must be first-order. The data was fit at multiple temperatures and the total moment was able to be extracted as a function of temperature, shown in Fig~\ref{figMomentvsTemp} to be approximately $4\mu_B$. Due to the low point density of the total moment as a function of temperature, it is difficult to fit the order parameter equation, but we have included a guide to the eye. The total moment found is lower than the saturated moment of $6.2\mu_B$ found from magnetization, but we know the saturated moment will be increased by the field-induced mixing of the low lying crystal field level.

\clearpage

\bibliographystyle{apsrev}

\end{document}